%% file: main.tex
\documentclass[10pt,letterpaper,twocolumn]{article}
\usepackage{manuscript}

\makeatletter
\AtBeginDocument{\immediate\write\@auxout{\string\citation{IEEEreferencecontrol}}}
\makeatother
\hypersetup{pdftitle={A Hidden Markov Analysis of Wet-Weather Tyre Crossover in Formula One},pdfauthor={Bo-Yi Lin and Michka Sharpe}}
\input{window_results}

\title{A Hidden Markov Analysis of Wet-Weather\\[2pt]Tyre Crossover in Formula One}
\author{\begin{minipage}{0.95\textwidth}\centering\input{author_block}\end{minipage}}
\date{}
\renewcommand{\shorttitle}{Wet-Weather Tyre Crossover}
\begin{document}
\twocolumn[\begin{@twocolumnfalse}
\maketitle

\begin{abstract}
Detecting a wet-weather regime and deciding when to replace slick tyres are distinct tasks. This study examines their relationship through a retrospective analysis of the 2024 British Grand Prix. A three-state hidden Markov model (HMM) represents normalized lap time and precipitation, while an entry rule requires both a high filtered Wet-state probability and sufficient pace deterioration. The analysis includes \WindowCarLaps{} car-laps and \WindowHMMFitN{} complete lap-bin observations within the available precipitation record. Each completed-lap signal takes effect on the following lap. The combined rule enters on lap \WindowEntryLap{}, matching the team-weighted modal reference; removing the pace or state condition advances entry to laps \WindowPosteriorEntry{} and \WindowPaceOnlyEntry{}, respectively. The combined entry is retained in all \WindowStableTeams{} team-exclusion refits, but refitting a specification with larger off-diagonal transition probabilities produces entry on lap \WindowStructureEarlyEntry{}. These comparisons illustrate how state inference can contextualize a performance criterion while exposing dependence on model assumptions. Because estimation and rule development use the same event, the correspondence establishes neither prospective accuracy nor a race-time advantage.
\end{abstract}

\vspace{10pt}
\end{@twocolumnfalse}]
\thispagestyle{plain}

\section{Introduction}
Wet-weather tyre selection requires a judgement about relative performance under changing track conditions. Rainfall provides environmental information, but the point at which intermediate tyres become preferable to slicks also depends on their response to those conditions. Pirelli describes a general crossover near 110--112\% of normal slick lap time \cite{pirelli2020crossover}. This guidance motivates a performance criterion; rainfall alone does not identify the crossover, and a temporary slowdown need not imply a persistent change.

Race-strategy research addresses this decision within broader models of tyre use, pit stops, and competition. Discrete-event and Monte Carlo simulations represent race dynamics and stochastic disturbances \cite{bekker2009strategy,heilmeier2020montecarlo}. Optimal-control and dynamic-programming formulations address tyre management and stop timing \cite{west2020tyre,carrasco2023pitstop}, while game-theoretic models include competitors' responses \cite{aguad2024game}. Supervised and reinforcement-learning approaches also select actions within race simulations \cite{heilmeier2020vse,thomas2026rsrl}. More directly, recurrent-network simulations have examined alternative wet-weather pit-stop timings \cite{mistry2025autof1}. These approaches evaluate actions through an assumed or estimated performance model. A related empirical question is whether evidence of a changing regime and evidence of pace deterioration support the same entry time.

Hidden Markov models provide a standard representation of sequential observations generated by unobserved states \cite{baum1966statistical,rabiner1989hmm}. State inference and action selection remain separate problems under partial observability \cite{kaelbling1998pomdp}: a high probability of a wet regime does not itself quantify the benefit of changing tyres. Here, we compare an HMM-and-pace rule with matched rules that remove either condition to reconstruct initial slick-to-intermediate entry within the precipitation record, and assess sensitivity to estimation and decision assumptions.

\section{Data and Evaluation Design}
\subsection{Observation window and preprocessing}
The supplied 2024 British Grand Prix export contains 960 car-laps from \WindowDrivers{} drivers. Lap timing, tyre compounds, pit events, and accuracy indicators are derived from FastF1 \cite{fastf1}. The external precipitation column is identified as five-minute accumulation in millimetres, consistent with the provider's parameter definition \cite{meteomatics_precip}. The original request, spatial coordinates, and measurement or model provenance are unavailable, so the precipitation series cannot be treated as a verified trackside measurement. FastF1's binary rainfall flag supplies a separate reporting reference and is not an HMM input.

The observation window runs from race start to the final recorded precipitation sample. Precipitation is matched backward to individual lap completions with a maximum age of 300\,s; measured zero rainfall remains valid. A car-lap is retained only if both its completion time and its original numbered-bin median completion time lie within the observation window. This gives laps 1--\WindowLastLap{}, \WindowCarLaps{} car-laps, and \WindowCleanLaps{} clean pace observations. Clean laps exclude the opening lap, pit-in and pit-out laps, deleted or generated laps, inaccurate records, and missing lap times. Exact alignment timestamps and input checksums are retained in the reproducibility manifest.

\begin{figure*}[!t]
\centering
\includegraphics[width=\textwidth]{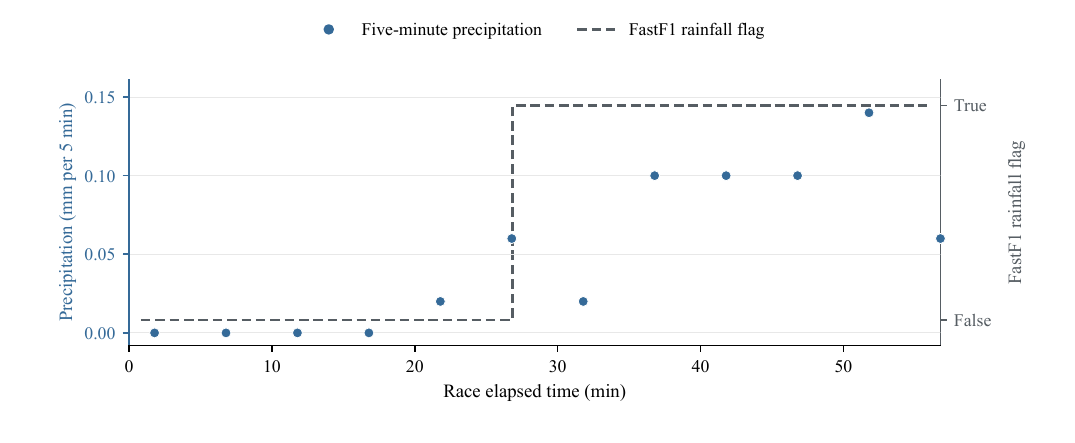}
\caption{Five-minute precipitation (blue points, left axis) and FastF1's rainfall flag (grey dashed step, right axis) over elapsed race time. The flag trace ends at its last report.}
\label{fig:weather}
\end{figure*}

\subsection{Pace reference and lap-bin observations}
For driver $i$, the dry reference $b_i$ is the median of the three fastest clean slick laps completed before the first positive precipitation report, with measured precipitation equal to zero. This calibration pool contains \WindowCalibrationCarLaps{} car-laps. The reference is held fixed throughout evaluation. Normalized lap time and its grid aggregate are
\begin{equation}
q_{i,t}=100\frac{L_{i,t}}{b_i},\qquad
g_t=\operatorname{median}_{i\in C_t}q_{i,t},
\label{eq:pace}
\end{equation}
where $L_{i,t}$ is lap time and $C_t$ contains drivers with clean observations in numbered lap bin $t$. Thus, $q_{i,t}=110$ denotes a lap time 10\% above that driver's dry reference. The precipitation input $r_t$ is the median of backward-matched values across retained car-laps in the bin. The HMM observes $\mathbf{x}_t=(g_t,r_t)^\top$; excluding the opening bin leaves \WindowHMMFitN{} consecutive complete observations.

The aggregate $g_t$ includes all observed tyre classes. It therefore describes realized grid performance and may change when drivers change tyres, even at the same underlying track condition. Applying the published slick-lap guidance to this aggregate is an operational extension, not a direct estimate of slick--intermediate performance equality. A supplementary comparison restricts the pace condition to clean slick laps while retaining the same HMM.

\subsection{Reference entry and action timing}
The observed reference is the first intermediate-tyre lap of a team-weighted consensus. Each active team contributes one vote when its available drivers agree on tyre class; split teams do not vote. The unique modal class defines the consensus, and ties yield no reference label. This summarizes observed choices rather than assuming that they minimize race time.

Evaluation begins after the calibration period, on laps \WindowEvalStart--\WindowLastLap{}. A signal formed from completed bin $t$ changes the recommendation on bin $t+1$. This convention distinguishes evidence from its effective action, but aggregation by numbered lap does not reproduce the asynchronous information available to each driver at a real pit decision. In addition, the HMM parameters are estimated retrospectively. The analysis therefore compares reconstructed entry times rather than claiming an online evaluation free of future information \cite{kaufman2012leakage}.

\Needspace{10\baselineskip}
\section{Hidden Markov Model and Entry Rule}
\subsection{State model and estimation}
Let $z_t\in\{D,M,W\}$ denote Dry, Damp, and Wet statistical regimes. Each feature is standardized using its mean and standard deviation over the fitted window. The emission model is
\begin{equation}
\tilde{\mathbf{x}}_t\mid z_t=k
\sim\mathcal{N}(\boldsymbol{\mu}_k,\boldsymbol{\Sigma}_k).
\label{eq:emission}
\end{equation}
The initial distribution and transition matrix are fixed:
\begin{equation}
\begin{gathered}
\boldsymbol{\pi}=(0.98,\,0.019,\,0.001),\\[3pt]
\mathbf{A}=
\begin{bmatrix}
0.92&0.08&0\\
0.10&0.80&0.10\\
0&0.12&0.88
\end{bmatrix}.
\end{gathered}
\label{eq:transition}
\end{equation}
Here, $A_{jk}=P(z_t=k\mid z_{t-1}=j)$. The matrix imposes persistent, adjacent-regime transitions and excludes direct Dry--Wet changes within one step. These are modelling assumptions, not estimated transition frequencies or physical constraints on rainfall. The three-state specification allows an intermediate regime between near-baseline and higher-rain, slower-pace observations.

Emission means and full covariance matrices are fitted by expectation--maximization \cite{dempster1977em}, using a modified covariance update in \texttt{hmmlearn} \cite{hmmlearn}. The model has 15 fitted emission parameters. Three deterministic marginal-quantile initializations are compared, and the highest-likelihood admissible fit is retained without team-consensus labels. Admissibility requires convergence, a monotonic likelihood, and increasing marginal pace and precipitation means in the fixed Dry--Damp--Wet state order. Each covariance update enforces $\boldsymbol{\Sigma}_k\succeq\epsilon\mathbf{I}$ with $\epsilon=10^{-3}$ in standardized coordinates; covariance priors are zero. Dry, Damp, and Wet label these fitted distributions rather than independently verified surface-water or grip states. Table~\ref{tab:emissions} reports marginal parameters on original scales; full covariance estimates and initialization diagnostics accompany the supplement.

\begin{table}[!t]
\caption{Estimated emission parameters on original scales}
\label{tab:emissions}
\centering\small
\input{tables/window_hmm_parameters}
\par\smallskip\footnotesize\raggedright
$g$: normalized lap time; $r$: five-minute precipitation accumulation; pp: percentage points of normalized lap time. The fitted covariance matrices are full; $\sigma$ denotes marginal standard deviation.
\end{table}

\subsection{Filtered state evidence and entry criterion}
Conditional on fitted parameters $\hat{\theta}$, the filtered probability is $\alpha_t(k)=P(z_t=k\mid\mathbf{x}_{2:t};\hat{\theta})$, with recursion
\begin{equation}
\alpha_t(k)=
\frac{f_k(\tilde{\mathbf{x}}_t)\sum_j A_{jk}\alpha_{t-1}(j)}
{\sum_\ell f_\ell(\tilde{\mathbf{x}}_t)\sum_j A_{j\ell}\alpha_{t-1}(j)},
\label{eq:filter}
\end{equation}
where $f_k$ is the Gaussian emission density. At the first complete bin, $\alpha_2(k)\propto\pi_k f_k(\tilde{\mathbf{x}}_2)$. The recursion preserves structural zeros and uses observations up to the current bin. However, $\hat{\theta}$ and the standardization parameters use the complete window, so these probabilities remain retrospective estimates.

For a post-calibration bin, define the qualifying indicator
\begin{equation}
u_t=\mathbf{1}\{r_t>0,\ \alpha_t(W)\geq\tau,\ g_t\geq\gamma\}.
\label{eq:criterion}
\end{equation}
Before evaluation or when required evidence is missing, $u_t=0$. The signal and effective entry times are
\begin{equation}
\begin{aligned}
T_{\mathrm{signal}}&=\min\left\{t:\prod_{h=0}^{m-1}u_{t-h}=1\right\},\\
T_{\mathrm{entry}}&=T_{\mathrm{signal}}+1.
\end{aligned}
\label{eq:entry}
\end{equation}
The $m$ bins must be consecutive; a missing bin resets confirmation. No qualifying sequence means no entry. Recommendations remain slick before $T_{\mathrm{entry}}$ and intermediate thereafter, restricting the task to the initial entry.

The default uses $\tau=0.5$, $\gamma=110$, and $m=1$, with $\gamma$ in the percentage units of Eq.~\eqref{eq:pace}. The explicit pace condition requires a specified level of deterioration even when precipitation dominates state assignment. Because pace also enters the emissions, the conditions use shared observations. The rule and model specification were developed while inspecting this event. They are not preregistered or independently validated decision settings.

\subsection{Comparisons and sensitivity analysis}
Matched component comparisons remove either $g_t\geq\gamma$ (posterior-only) or $\alpha_t(W)\geq\tau$ (pace-only), retaining positive precipitation, confirmation length, and next-bin action timing. A non-dry comparator uses $\alpha_t(M)+\alpha_t(W)\geq0.5$ without a pace condition. An always-slick comparator makes no entry. Each method retains the intermediate recommendation after entry. The primary descriptive measure is entry error, $\Delta T=T_{\mathrm{entry}}-T_{\mathrm{reference}}$.

Decision sensitivity crosses seven posterior thresholds from 0.40 to 0.95, pace thresholds of 110\%, 111\%, and 112\%, and one- or two-bin confirmation, holding the fitted HMM fixed. Team-exclusion analyses remove each team's pace observations and refit standardization and emissions from all three initializations, keeping weather and the lap-bin clock fixed. Model sensitivity crosses covariance floors of $10^{-4}$, $10^{-3}$, and $10^{-2}$ with off-diagonal transition-probability multipliers $\kappa\in\{0.5,1,1.5\}$. For $j\ne k$, $A^{(\kappa)}_{jk}=\kappa A_{jk}$, and each diagonal is adjusted to retain a unit row sum. The graph and initial probabilities remain fixed, and each specification is refitted. A further check uses clean slick-lap medians for the pace condition. All are within-event sensitivity analyses, not independent race replications.

\section{Results}
\subsection{Regime evidence and pace deterioration}
The precipitation record became positive before the FastF1 rainfall flag (Fig.~\ref{fig:weather}). Differences in sampling and spatial provenance preclude interpreting this as a validated lead in track-wetness detection. The model instead combined precipitation with observed pace. Its Dry, Damp, and Wet mean normalized lap times were approximately 100.3\%, 103.0\%, and 110.8\%, respectively (Table~\ref{tab:emissions}).

\begin{figure*}[!t]
\centering
\includegraphics[width=\textwidth]{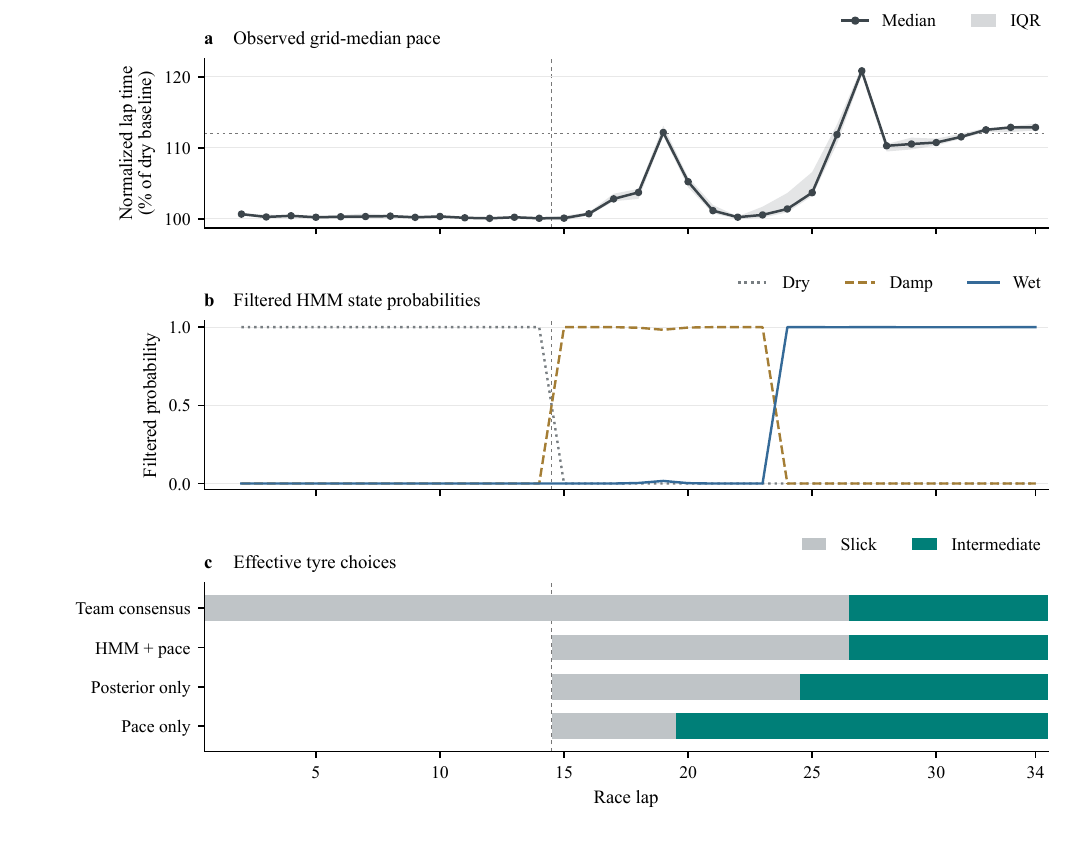}
\caption{Pace, state evidence, and effective tyre recommendations. (a) Median normalized lap time with shaded within-bin interquartile range across clean car-laps; reference lines mark 110\% and 112\%. (b) Retrospective filtered state probabilities with a 0.5 reference line. (c) Team consensus and entry rules. The vertical dashed line marks evaluation from lap \WindowEvalStart{}.}
\label{fig:crossover}
\end{figure*}

The two components supported different entry signals (Fig.~\ref{fig:crossover}). The Wet posterior first exceeded 0.5 on lap \WindowPosteriorSignal{}, when the grid median was 101.4\%, resulting in posterior-only entry on lap \WindowPosteriorEntry{}. The pace condition withheld combined entry until the lap-\WindowPrimarySignal{} median reached \WindowSignalPacePct{}\%. That signal used \WindowSignalCleanCars{} clean car-laps and produced effective entry on lap \WindowEntryLap{}, coinciding with the observed consensus (Table~\ref{tab:benchmark}). The pace-only rule had already signalled after the lap-\WindowPaceOnlySignal{} slowdown, producing entry on lap \WindowPaceOnlyEntry{}. The non-dry comparator entered on lap \WindowNonDryEntry{}.

\begin{table}[!t]
\caption{Entry timing on the common evaluation window}
\label{tab:benchmark}
\centering\small
\input{tables/window_benchmark}
\par\smallskip\footnotesize\raggedright
Signal: completed bin supplying the decision. Entry: next effective bin. $\Delta T$: entry minus the observed consensus entry, lap \WindowReferenceEntry{}, measured in laps. A dash denotes no entry.
\end{table}

The combined rule excluded both early component signals in this event. Agreement across the \WindowEvalLaps{} evaluated bins follows from a single matched entry and is not an independent predictive success rate. Clean-bin support also varied during the pit cycle, falling to two car-laps on lap 27; the trajectory should therefore be read with its changing observational support.

\subsection{Dependence on estimation and decision assumptions}
The selected emission fit converged monotonically in seven iterations. On the standardized feature scale, two starts reached a log likelihood of 26.899, whereas the remaining start reached 3.838. The lower-likelihood fit produced posterior-only entry two laps later, while combined entry was unchanged. Thus, an individual initialization would have changed the apparent component comparison. Complete fits and regularization diagnostics are reported in Supplementary Section~S1.

With one-bin confirmation, all tested posterior thresholds retained lap-27 entry at pace thresholds of 110\% and 111\%; 112\% moved entry to lap 28. Two-bin confirmation gave entry on lap 28 at 110--111\%, but on lap 34 at 112\%, because a temporary pace recovery interrupted the qualifying sequence. All \WindowStableTeams{} team-exclusion refits retained combined entry on lap 27. The slick-only pace condition also gave lap-27 entry, although evidence was available in only \WindowSlickEvalAvailable{} of the \WindowEvalLaps{} evaluation bins. The remaining \WindowSlickEvalMissing{} bins lacked a complete slick-only signal and were not imputed.

Changing the transition specification and refitting the emissions materially altered the entry. At $\kappa=0.5$ and $\kappa=1$, all three covariance floors retained combined entry on lap 27 and posterior-only entry on lap \WindowPosteriorEntry{}. At $\kappa=1.5$, both rules entered on lap \WindowStructureEarlyEntry{} for all floors. Hence, stability across the decision grid and team exclusions did not extend to the assumed regime-transition probabilities. The nine model specifications and complete decision sensitivities are reported in Supplementary Section~S2. Near-saturated fitted probabilities also limit the interpretation of posterior-threshold stability as evidence of calibration.

\section{Discussion}
Under the default model, the initial slowdown preceded the Wet-state signal, which emerged after pace had returned near the dry reference. Combined entry required renewed pace deterioration. Under the specification with larger off-diagonal transition probabilities and refitted emissions, Wet-state evidence accompanied the initial slowdown, and combined entry advanced. This separates regime inference from a performance threshold while showing that their timing correspondence is conditional on model assumptions.

The entry rule could supply a trigger for a strategy model. Race-strategy studies evaluate tyre choices and pit-stop timing using performance and race-dynamics models \cite{bekker2009strategy,carrasco2023pitstop,aguad2024game}. Our observations do not identify counterfactual tyre pace, pit traffic, or rivals' responses, so agreement with team choices does not establish a race-time saving. Evaluating an alternative action requires a validated action-dependent simulator or off-policy identification assumptions \cite{jiang2016offpolicy}.

The main inferential constraint is estimation and rule development on one event. Fifteen emission parameters are fitted to \WindowHMMFitN{} bins, which reuse only \WindowCleanRainSamples{} distinct precipitation timestamps. Near-zero dry precipitation and the covariance bound contribute to sharply separated probabilities; Gaussian emissions are a phenomenological approximation for non-negative precipitation, not a physical rainfall model. Team exclusions reuse the same rain sequence and event development, assessing dependence on pace contributors rather than generalization across races. Available weather metadata also prevent independent verification of track-level exposure.

Pace introduces a second constraint. Fixed dry references do not remove fuel, tyre wear, traffic, or changing tyre composition, and the clean sample becomes small during pit cycles. Bayesian state-space tyre models explicitly separate latent pace from fuel effects \cite{cappello2026statespace}; comparable decomposition is absent here. Because tyre choices affect pace observations, the team-choice reference is not independent of the HMM inputs and does not measure track grip. Out-of-sample assessment is needed \cite{tashman2000forecast}. For this task, a prospective evaluation could fix model and rule settings using separate development events, then test entry timing with per-car availability timestamps. A complementary tyre-performance model would be required to test whether the resulting entry improves race outcomes.

\section{Conclusion}
Under the default specification, an HMM-and-pace criterion reconstructed the observed intermediate entry, whereas either component acted earlier. The transition sensitivity shows that this correspondence depends on the state model. Independent-event evaluation and a model of tyre-change consequences are required to assess prospective strategy value.

\bibliographystyle{IEEEtran}
\bibliography{references}
\end{document}


\twocolumn[\begin{@twocolumnfalse}
\maketitle
\end{@twocolumnfalse}]
\thispagestyle{plain}
\renewcommand{\thefigure}{S\arabic{figure}}
\renewcommand{\thetable}{S\arabic{table}}

\section*{S1. Observations and estimation}
The window retained \WindowCarLaps{} car-laps from \WindowDrivers{} drivers through lap \WindowLastLap{}, including \WindowCleanLaps{} clean observations. Clean laps used \WindowCleanRainSamples{} of \WindowRainSamples{} precipitation timestamps; reuse across bins does not add independent weather measurements. Fixed driver references used the three fastest eligible slick laps from the \WindowCalibrationCarLaps{} pre-rain, zero-precipitation calibration laps. Evaluation covered laps \WindowEvalStart--\WindowLastLap{}; pit-in and pit-out laps informed tyre occupancy but not pace.

The hidden Markov model (HMM) fitted 15 emission parameters to \WindowHMMFitN{} consecutive complete bins of standardized pace and precipitation, with fixed initial and transition probabilities. Each start allowed 250 iterations with tolerance $10^{-4}$, zero covariance prior and eigenvalue floor $\epsilon=10^{-3}$ in standardized coordinates. Admissible fits required monotonic convergence and increasing pace and precipitation means in Dry--Damp--Wet order.

\begin{table}[ht]
\centering\small
\caption{Initialization fits; log likelihoods use standardized features.}\label{tab:starts}
\input{tables/window_initializations}
\end{table}

All three starts converged (Table~\ref{tab:starts}). The quartile start had the highest tested likelihood, without guaranteeing a global optimum. Combined entry was lap \WindowEntryLap{} for every start; posterior-only entry was lap 27 for the lower-likelihood fit and lap 25 otherwise.

In original $(g,r)$ coordinates, the symmetric covariance entries $(\Sigma_{gg},\Sigma_{gr},\Sigma_{rr})$ were $(0.0324095,0,2.30376\times10^{-6})$ for Dry, $(13.3559,0.0543428,3.55649\times10^{-4})$ for Damp and $(23.0672,0.0148745,2.38675\times10^{-4})$ for Wet. The Dry covariance reached the floor in both standardized directions; its zero precipitation mean challenges the Gaussian approximation. State probabilities lack independent grip calibration.

\section*{S2. Decision and model sensitivity}
Seven posterior thresholds (0.40, 0.50, 0.60, 0.70, 0.80, 0.90, 0.95), three pace thresholds and one- or two-bin confirmation gave \WindowSensitivityN{} settings. Entry depended on pace and confirmation, but not posterior threshold (Table~\ref{tab:sensitivity}). At 112\%, pace recovery interrupted two-bin confirmation, delaying entry to lap 34. Only two clean car-laps supported the lap-27 pace estimate.

\begin{table}[ht]
\centering\small
\caption{Entry laps across posterior thresholds 0.40--0.95.}\label{tab:sensitivity}
\input{tables/window_sensitivity}
\end{table}

Removing each team's pace observations and refitting scaling and emissions from three starts gave 30 convergent fits. Weather and bin-completion times remained fixed. All \WindowStableTeams{} refits retained combined entry on lap \WindowEntryLap{} and posterior-only entry on lap \WindowPosteriorEntry{}. These perturbations share one event.

The structural grid crossed covariance floors $\epsilon\in\{10^{-4},10^{-3},10^{-2}\}$ with transition multipliers $\kappa\in\{0.5,1,1.5\}$. Off-diagonal probabilities became $A^{(\kappa)}_{jk}=\kappa A_{jk}$; diagonals retained unit row sums. Structural zeros and initial probabilities were fixed. All 27 fits from three starts converged with admissible marginal orderings. Increasing $\kappa$ to 1.5 advanced both entries to lap \WindowStructureEarlyEntry{} (Table~\ref{tab:structure}), showing dependence on transition assumptions. Likelihoods across different regularization constraints were not used to rank validated models.

\begin{table}[ht]
\centering\small
\caption{Entry laps after transition and covariance refits. Combined: HMM and pace; posterior-only: Wet-state rule. Reference: lap \WindowReferenceEntry{}.}\label{tab:structure}
\input{tables/window_structure_sensitivity}
\end{table}

\section*{S3. Slick-only evidence and timing checks}
Replacing only the pace condition with clean slick-lap medians retained entry on lap \WindowSlickEntryLap{}. Evidence was available in \WindowSlickEvalAvailable{} of \WindowEvalLaps{} evaluation bins; the remaining \WindowSlickEvalMissing{} were not imputed. Recommendations remained intermediate after entry.

Signals from bin $t$ applied at $t+1$; missing evidence or bin gaps reset confirmation, and internal missing HMM emissions were rejected. Numerical checks covered likelihood selection, covariance bounds, state ordering, probability normalization and forbidden transitions. Fixed-parameter filtering was checked for future-observation dependence; full-window parameter estimation nevertheless makes inference retrospective.

\clearpage
\twocolumn[{\begin{@twocolumnfalse}
\vspace*{10pt}
\section*{S4. Individual tyre trajectories}
Figure~\ref{fig:drivers} shows retained tyre observations by team; gaps indicate missing observations. On lap \WindowReferenceEntry{}, four teams voted intermediate and three voted slick; three split teams abstained. This is a modal observed choice, not unanimity or evidence of a race-time-optimal action.

\begin{minipage}{\textwidth}
\centering\includegraphics[width=\textwidth]{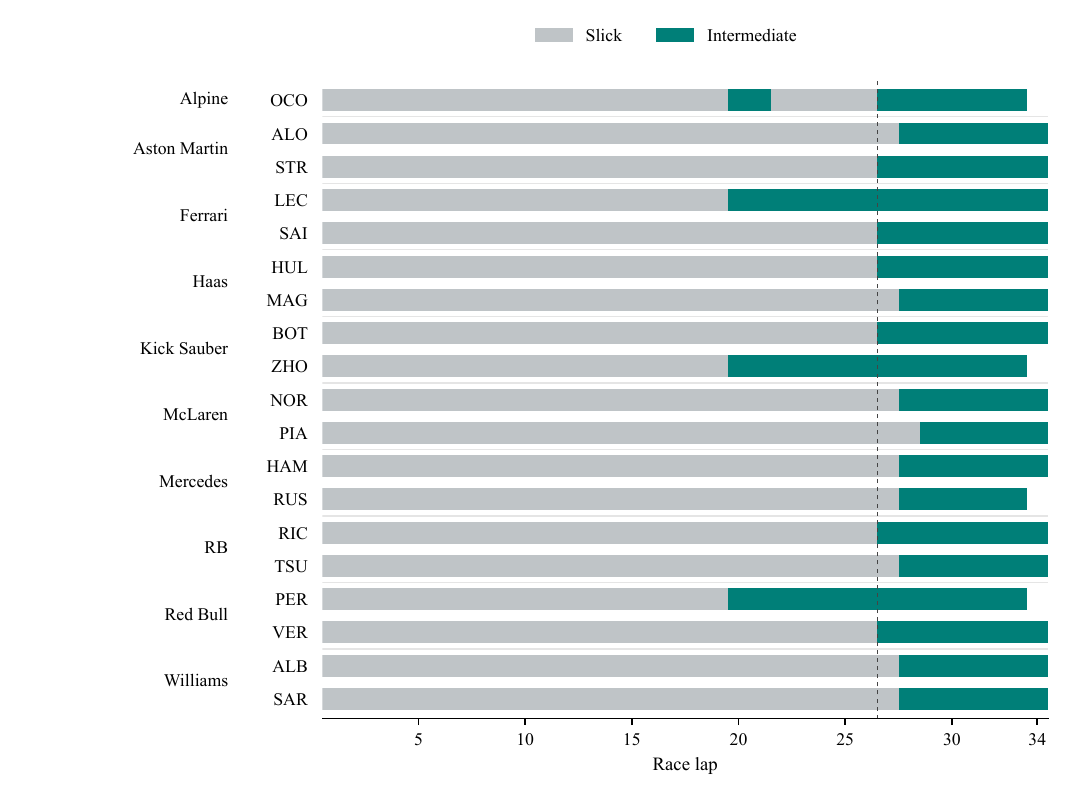}
\captionsetup{hypcap=false}
\captionof{figure}{Observed tyre classes through lap \WindowLastLap{}, grouped by team. The dashed line marks reference entry on lap \WindowReferenceEntry{}; blank intervals indicate missing observations.}\label{fig:drivers}
\end{minipage}
\end{@twocolumnfalse}}]

%% file: window_results.tex
\newcommand{\WindowLastLap}{34}
\newcommand{\WindowCarLaps}{642}
\newcommand{\WindowCleanLaps}{565}

\newcommand{\WindowEvalStart}{15}
\newcommand{\WindowEvalLaps}{20}
\newcommand{\WindowEntryLap}{27}

\newcommand{\WindowStableTeams}{10}
\newcommand{\WindowHMMFitN}{33}

\newcommand{\WindowCalibrationCarLaps}{236}
\newcommand{\WindowDrivers}{19}

\newcommand{\WindowCleanRainSamples}{11}
\newcommand{\WindowReferenceEntry}{27}
\newcommand{\WindowSignalPacePct}{111.9}
\newcommand{\WindowSignalCleanCars}{11}

\newcommand{\WindowStructureEarlyEntry}{20}

\newcommand{\WindowSlickEvalAvailable}{13}
\newcommand{\WindowSlickEvalMissing}{7}

\newcommand{\WindowPrimarySignal}{26}
\newcommand{\WindowPosteriorEntry}{25}
\newcommand{\WindowPosteriorSignal}{24}
\newcommand{\WindowPaceOnlyEntry}{20}
\newcommand{\WindowPaceOnlySignal}{19}
\newcommand{\WindowNonDryEntry}{16}

%% file: author_block.tex
{\large\bfseries Bo-Yi Lin\textsuperscript{1,2}, Michka Sharpe\textsuperscript{2}\par}
\vspace{4pt}
{\normalsize\mdseries \textsuperscript{1}\,Cranfield University\par
\textsuperscript{2}\,\href{https://www.codefxmotorsports.com/}{CodeF(x) LLC}\par}
\vspace{3pt}
{\small\mdseries Email: \href{mailto:boki1027uk@gmail.com}{boki1027uk@gmail.com}\par}

%% file: tables/window_hmm_parameters.tex
\begin{tabular}{lrrrr}
\toprule
State & $\mu_g$ (\%) & $\mu_r$ (mm) & $\sigma_g$ (pp) & $\sigma_r$ (mm) \\
\midrule
Dry & 100.27 & 0.000 & 0.18 & 0.002 \\
Damp & 102.95 & 0.033 & 3.65 & 0.019 \\
Wet & 110.83 & 0.107 & 4.80 & 0.015 \\
\bottomrule
\end{tabular}

%% file: tables/window_benchmark.tex
\begin{tabular}{lrrr}
\toprule
Method & Signal & Entry & $\Delta T$ \\
\midrule
Non-dry HMM & 15 & 16 & -11 \\
HMM + pace & 26 & 27 & +0 \\
Wet-state HMM & 24 & 25 & -2 \\
Pace only (110\%) & 19 & 20 & -7 \\
Always slick & --- & --- & --- \\
\bottomrule
\end{tabular}

%% file: tables/window_initializations.tex
\begin{tabular}{lrr}
\toprule
Initialization quantiles & Iterations & Final log likelihood \\
\midrule
0.15 / 0.50 / 0.85 & 12 & 3.838471 \\
0.25 / 0.50 / 0.75 & 7 & 26.898697 \\
0.10 / 0.30 / 0.80 & 6 & 26.898697 \\
\bottomrule
\end{tabular}

%% file: tables/window_sensitivity.tex
\begin{tabular}{lrr}
\toprule
Pace threshold $\gamma$ (\%) & $m=1$ & $m=2$ \\
\midrule
110 & 27 & 28 \\
111 & 27 & 28 \\
112 & 28 & 34 \\
\bottomrule
\end{tabular}

%% file: tables/window_structure_sensitivity.tex
\begin{tabular}{lrrr}
\toprule
$\kappa$ & $\epsilon$ & Combined & Posterior-only \\
\midrule
0.5 & 0.0001 & 27 & 25 \\
0.5 & 0.0010 & 27 & 25 \\
0.5 & 0.0100 & 27 & 25 \\
1.0 & 0.0001 & 27 & 25 \\
1.0 & 0.0010 & 27 & 25 \\
1.0 & 0.0100 & 27 & 25 \\
1.5 & 0.0001 & 20 & 20 \\
1.5 & 0.0010 & 20 & 20 \\
1.5 & 0.0100 & 20 & 20 \\
\bottomrule
\end{tabular}